\documentclass[10pt,a4in]{article}
\usepackage{graphicx}
\usepackage{geometry}
\usepackage[francais,english]{babel}
\begin{document}
\setcounter{page}{0}
\thispagestyle{empty}

\renewcommand{\thesection}{\Roman{section}}
\begin{center}
\large{\bf Numerical approach to the lowest bound state of muonic three-body systems}\\
\vspace{0.5cm}
\normalsize\rm
MD. ABDUL KHAN\\
\footnotesize
Assistant Professor, Department of Physics, Aliah University, \\ Action Area- II, Plot No. IIA/27, Newtown, Kolkata-700156, India\\ 
{\it Email:} $  drakhan.rsm.phys@gmail.com; drakhan.phys@aliah.ac.in$\\
\end{center}
\vspace{1.0cm}
\rm
%\tableofcontents
\begin{abstract}
In this paper, calculated energies of the lowest bound state of Coulomb three-body systems containing an electron ($e^-$), a negatively charged muon ($\mu^-$) and a nucleus ($N^{Z+}$) of charge number Z are reported. The 3-body relative wave function in the resulting Schr\"odinger equation is expanded in the complete set of hyperspherical harmonics (HH). Use of the orthonormality of HH leads to an infinite set of coupled differential equations (CDE) which are solved numerically to get the energy E.
\end{abstract}
\hspace{1.0cm}{\it Keywords:} Hyperspherical Harmonics (HH), Raynal-Revai Coefficient (RRC), \\\hspace*{2.5cm}Renormalized Numerov Method (RNM), Exotic Ions.\\ \hspace*{1.0cm}{\it PACS:} 02.70.-c, 31.15.-Ar, 
31.15.{\cal J}a, 36.10.{\cal E}e.

\section{Introduction}
Atoms and ions containing exotic particles like muon, kaon, taon, baryon and their antimatters are of immense have become an interseting research topic in many branches of physics including atomic, nuclear and elementary particle physics, plasma and astrophysics, experimental physics [1-2]. Out of the many probable species, the muonic helium atom ($^4$He$^{2+}\mu^-e^-$) formed by replacing an orbital electron of neutral helium is the simplest one which can be treated as a three-body atomic system [3] was first formed and detected by Souder et al [4]. Muon being about 207 times heavier than an electron, the size of the muonic helium atom is smaller by a factor of about 1/400 of the ordinary electronic helium atom. Similar arguments hold for ($^3$He$^{2+}\mu^-e^-$) also. Several reasons have been stated in the literature regarding the importance of these exotic atoms: 
i) Muonic helium atoms are the unusual pure atomic three body systems without any restriction due to Pauli exclusion principle for electron and muon being non-identical fermions.
ii) These are the by-products of the process of muon catalyzed fusion and study of these may yield useful information to understand the fusion reactions properly [5-6].
iii) The electromagnetic interaction between the electron and negatively charged muon can be better understood by this simplest muonic system by precise measurements of hyperfine structure [7-8] of the ground-state, in terms of interaction between the spin magnetic moments of muon and electron.

As exotic particles are mostly unstable, their parent atoms (or ions) are also very short lived. These exotic short-lived atoms or ions can be formed by trapping the accelerated exotic particles inside matter and replacing one or more electron(s) in an ordinary atom by exotic particle(s). The absorbed exotic particle revolves round the nucleus of the target atom in orbit of radius equal to that of the electron before its ejection from the atom. Which subsequently cascades down the ladder of resulting exotic atomic states by the  emission of X-rays and Auger transitions before being lost on its way to the nucleus. If the absorbed exotic particle is a negatively charged muon, it passes through various intermediate atmospheres before being trapped in the vicinity of the atomic nucleus [9]. 
In the course of its journey inside the matter, it scatters from atom to atom as free electron and gradually loses its energy until it is captured into an atomic orbit. In the lowest energy level (1S), it experiences only Coulomb interaction with nuclear protons while it experiences weak interaction with the rest of the nucleons. On the other hand if the exotic particle is a hadron like a kaon, pion or anti-proton, the cascade ends earlier in all exotic atoms except the lighter ones with atomic number 1 or 2, due to nuclear absorption or annihilation of the particle by the short-range strong interaction. As discussed above, exotic atoms (or ions) are produced by replacing one or more electron(s) of neutral atoms by one or more exotic particle(s) like muon, pion, kaon, anti-proton having an electric charge equal to that of the electron [10]. The most studied exotic few-body Coulomb system are the muonic atoms (or muonic ions) which are formed by removing one or more orbital electron(s) by one or more negatively charged muon(s). However the present communication we shall consider only those atoms or ions in which the positively charged nucleus is being orbited by one electron and one negatively charged muon. 
By far exotic muonic atoms were widely used to probe a number of atomic properties including nature and strength of eletron-muon interaction [11], these were considered an effective testing probe to study the electromagnetic properties of nuclei [12]. A number of observables like magnetic hyperfine structure by Johnson and Sorensen [13], isotopic shifts in muonic spectra of isotopes of the chemical elements like Ca, Cr, Cu, Mo etc  by Macagno et al [14], perturbation calculation for hyperfine structure of muonic helium atom ($\mu e ^4$He), Lamb-shift in the muonic deuterium ($\mu$D) by Krutov and Martynenko [15] have been studied. Frolov  studied bound-state properties of $^3$He$^{2+}μ^−e^−$ and $^4$He$^{2+}μ^−e^−$ [16] and beryllium-muonic ions [17-18] with high accuracies. Flambaum [19] in 2008 investigated the effect of bound muons, pions, kaons etc on the fission barrier and stability of highly charged nuclei. In addition to ongoing projects many new experiments have been proposed in Muon Science Laboratory, RIKEN [20]. Some physical aspects of the reaction and dynamics of muonic helium atom as a classical three-body problem have been described by Sutchi et al [21]. Experimental investigation on the reactions of muonic helium and muonium with H$_2$ by Donald G Fleming and Co-workers [22] have been reported in the literature, although there may be more reported works in this direction.

Out of several theoretical methods applied to study the bound state properties of the atomic few-boy systems some may include integral differential approach by Sultanov et al [23],  potential harmonic approximation approach by Yalcin et al [24], Faddeev approach by Dodd [25], angular correlated configuration interaction (ACCI) approach by Rodriguez et al [26], Smith Jr et al [27], variational expansion approach by Frolov [16,18,28-31], and by Frolov et al [32-36]. Variational calculation for muonic few-body systems was initiated in the sixties by Halpern [37], Carter [38-39] and Delves et al [40]. Later in the late eighties Kamimura [41] used variational method for the bound D- state in $dtu$. In the early eighties Vinitsky {\it et al} [42] used non-variational approach for the muonic molecular ions. In the last decade of the twentieth century, Krivec and Mandelzweig [43] performed a non-variational precision calculation for Muonic helium atom ($^4$H$e^{2+}\mu^-e^-$) employing the Correlated Function HH method.

We adopt hyperspherical harmonics expansion approach to the ground state of atoms/ ions containing an orbital electron plus a negatively charged muon revolving round the positively charged nucleus thereby forming a three-body system. We assume the model to be valid for the electromagnetic interaction of the valence particles with the nucleus which is sufficiently weak to influence the internal structure of the nucleus. Again, the fact that the muon is much lighter than the nucleus allows us to regard the nucleus to remain a static source of electrostatic interaction. However, a hydrogen-like two-body model, consisting a quasi-nucleus ($\mu^-N^{(Z-2)+}$, formed by the muon and the nucleus) plus an orbital electron can also be tested. This is because the muon being about 200 times heavier than an electron, has an orbital radius of about 1/200 times that of an orbital electron. Hence, there is a fair possibility of forming the said quasi-nucleus. 

In HHE formalism, for a general three-body system consisting three unequal mass particles the choice of Jacobi coordinates correspond to three different partitions and in the $i^{th}$ partition, particle labeled by $\lq i$' remains as a spectator while the remaining two particles labeled $\lq j$' and $\lq k$' form the interaction pair. Thus the total potential contains three binary interaction terms (i.e. $V = V_{jk}(r_{jk}) + V_{ki}(r_{ki}) + V_{ij}(r_{ij})$) and for computation of matrix element of V($r_{ij}$), potential of the $(ij)$ pair, the chosen HH is expanded in the set of HH corresponding to the partition in which $\vec{r_{ij}}$ is proportional to the first Jacobi vector [44] by the use of Raynal-Revai coefficients (RRC) [45]. The energies obtained  for the lowest bound S-state is compared with the ones of the literature. 

In Section II, we briefly introduce the hyperspherical coordinates and the scheme of the transformation of HH belonging  to two different partitions. Results of calculation and discussions will be presented in Section III and finally we shall draw our conclusion in section IV.

\section{HHE Method}
The choice of Jacobi coordinates for systems of three particles of mass $m_{i}$, $m_{j}$, $m_{k}$ is shown in Fig.1.
\begin{figure}
\centering
\fbox{\includegraphics[width=0.75\linewidth, height=0.75\linewidth]{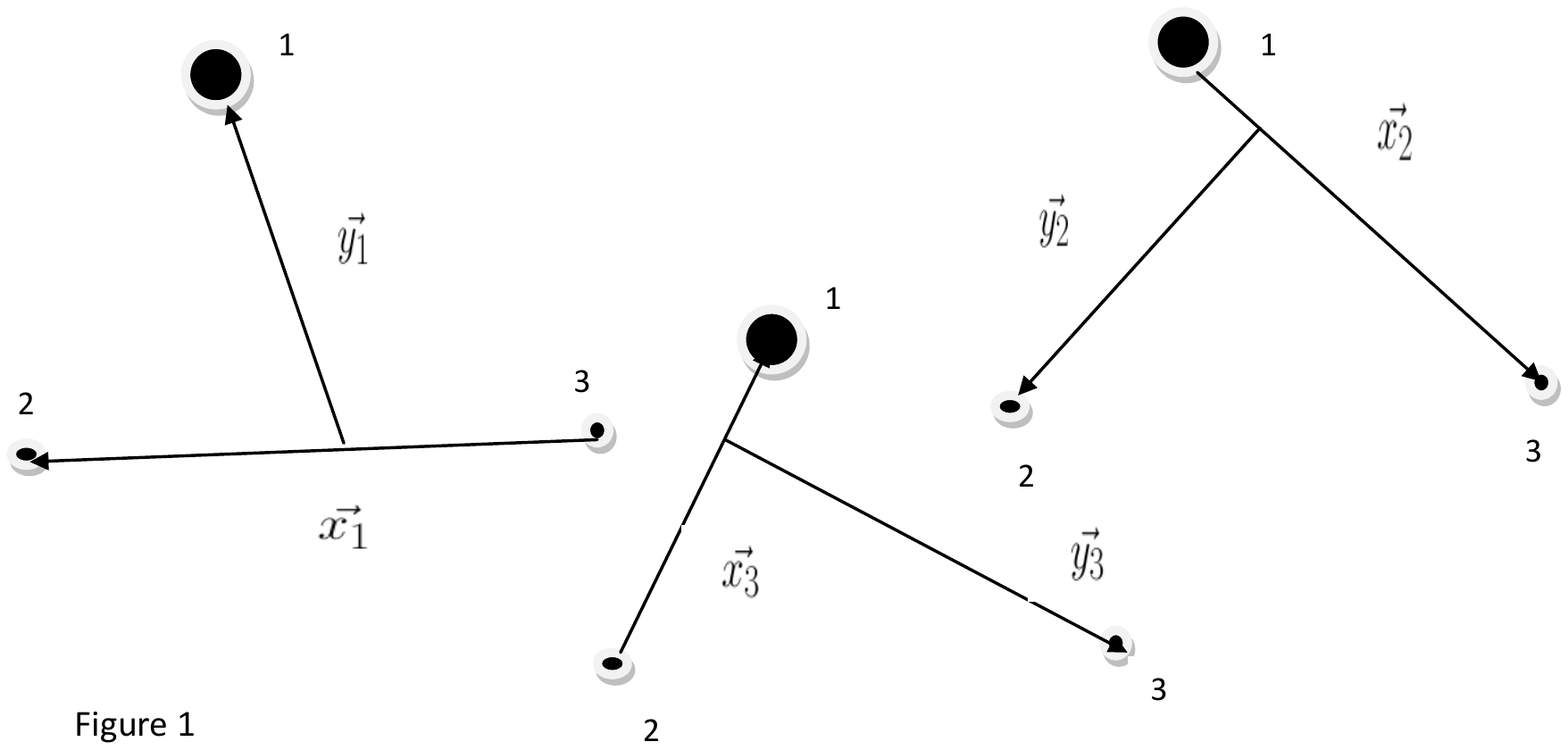}}
\caption{Choice of Jacobi coordinates in different partitions of a three-body system.}
\label{fig:boxed_graphic}
\end{figure}
The Jacobi coordinates [46] in the $i^{th}$ partition can be defined as: 
\begin{equation}
 \left.  \begin{array}{ccl}
   \vec{x_{i}} & = & \left[ \frac{m_{j} m_{k}M}{m_{i}(m_{j}+m_{k})^{2}}
\right] 
^{\frac{1}{4}} (\vec{r_{j}} - \vec{r_{k}}) \\ 
   \vec{y_{i}} & = & \left[ \frac{m_{i} (m_{j}+m_{k})^{2}}
{m_{j} m_{k} M}\right]^{\frac{1}{4}} \left( \vec{r_{i}} - \frac{m_{j}
\vec{r_{j}} + m_{k} \vec{r_{k}}}{ m_{j} + m_{k}} \right) \\
\vec{R} & = & (m_i\vec{r_i}+ m_j\vec{r_j} + m_k\vec{r_k})/M
    \end{array}  \right\} 
\end{equation}
where $M=m_i+m_j+m_k$ and the sign of $\vec{x_{i}}$ is determined by the condition that ($ i, j, k $) should form a cyclic permutation of (1, 2, 3).\\
The Jacobi coordinates are connected to the hyperspherical coordinates [47] as
\begin{equation}
 \left. \begin{array}{cclcccl}
 x_{i} & = & \rho \cos \phi_{i}&;& y_{i} & = & \rho \sin \phi_{i}\\
\rho&=& \sqrt{ x_{i}^{2} + y_{i}^{2}}&;&\phi_i&=&tan^{-1}(\eta_i/\Psi_i)
   \end{array} \right\} 
\end{equation}
The relative three-body Schr\H{o}dinger's equation in hyperspherical coordinates can be written as
\begin{equation}
\left[ - \frac{\hbar^{2}}{2\mu}\left\{ \frac{\partial^2}{\partial\rho^2}+ \frac{5}{\rho} \frac{\partial}{\partial\rho}+
\frac{\hat{{K}}^{2}(\Omega_{i})}{\rho^{2}} \right\}+ V (\rho, \Omega_{i}
) - E \right] 
\Psi (\rho, \Omega_{i} ) \:=\: 0
\end{equation}
where $\Omega_{i} \rightarrow \{ \phi_{i}, \theta_{x_{i}}, \phi_{x_{i}},
\theta_{y_{i}}, \phi_{y_{i}} \}$, effective mass ${\mu=\left[\frac{m_{i} m_{j} m_{k}}{M}\right]^{\frac{1}{2}}}$, potential $V(\rho, \Omega_{i})$ = $ V_{jk} + V_{ki} + V_{ij}$. The square of hyper angular momentum operator $\hat{K}^{2}(\Omega_{i})$ satisfies the eigenvalue equation [47]
\begin{equation}
\hat{{K}}^{2}(\Omega_{i}) {\cal Y}_{K \alpha_{i}}(\Omega_{i})=K
(K + 4 )  
{\cal Y}_{K \alpha_{i}}(\Omega_{i})
\end{equation}
where the eigen function ${\cal Y}_{K \alpha_{i}}(\Omega_{i})$ is the hyperspherical harmonics (HH). An explicit expression for the HH with specified grand orbital angular momentum $L(=\mid\vec{l_{x_i}} + \vec{l_{y_i}}\mid)$ and its projection $M$ is given by
\begin{equation}
 \begin{array}{rcl}
{\cal Y}_{K \alpha_{i}}(\Omega_{i})& \equiv &
{\cal Y}_{K l_{x_{i}} l_{y_{i}} L M}(\phi_{i}, \theta_{x_{i}},
\phi_{x_{i}}, \theta_{y_{i}}, \phi_{y_{i}})\\
 & \equiv & ^{(2)}P_{K}^{l_{x_{i}} l_{y_{i}}}(\phi_{i})
\left[Y_{l_{x_i}}^{m_{x_i}}(\theta_{x_{i}}, \phi_{x_{i}}) Y_{l_{y_{i}}}^{
m_{y_{i}}}(\theta_{y_{i}}, \phi_{y_{i}}) \right]_{L M}
 \end{array}
\end{equation}
with $\alpha_{i}\equiv \{l_{x_{i}}, l_{y_{i}}, L, M\}$ and $[ ]_{L M}$ denoting angular momentum coupling.
The hyper-angular momentum quantum number $K$($=2n_i+ l_{x_{i}} + l_{y_{i}}$; $n_i\rightarrow$ a non-negative integer) is not a conserved quantity for the three-body system.
In a given partition (say partition $\lq\lq i$"), the  wave-function $\Psi(\rho, \Omega_{i})$ is expanded in the complete set of HH 
\begin{equation}
\Psi(\rho, \Omega_{i}) = \sum_{K\alpha_{i}}\rho^{-5/2}U_{K\alpha_{i}}   
  (\rho) {\cal Y}_{K\alpha_{i}}(\Omega_{i})
\end{equation}
Substitution of Eq. (6) in Eq. (3) and use of ortho-normality of HH, leads to a set of coupled
differential equations (CDE) in $\rho$
\begin{equation}
\begin{array}{cl}
& \left[ -\frac{\hbar^{2}}{2\mu} \frac{d^{2}}{d\rho^{2}}
+\frac{(K+3/2)(K+5/2) \hbar^2}{2\mu\rho^{2}} - E \right] 
U_{K \alpha_{i}}(\rho)  \\
+ & \sum_{K^{\prime} \alpha_{i}^{\prime}} <K \alpha_{i}
\mid V(\rho, \Omega_{i}) \mid K^{\prime} \alpha_{i}^{\prime}
>U_{K^{\prime} \alpha_{i}^{~\prime}}(\rho) \: = \: 0.
\end{array}
\end{equation}
where
\begin{equation}
<K \alpha_{i} |V(\rho,\Omega_i)| K^{\prime}, \alpha_{i}^{\prime}> = \int
{\cal Y}_{K\alpha_{i}}^{*}(\Omega_{i}) V(\rho, \Omega_{i}) {\cal
Y}_{K^{\prime} 
 \alpha_{i}^{~\prime}}(\Omega_{i}) d\Omega_{i}
\end{equation}
   Calculation of the matrix elements of the form $<{\cal Y}_{K \alpha_{i}}(\Omega_{i}) \mid V_{j k}(x_{i}) \mid {\cal Y}_{K^{\prime}  \alpha_{i}^{\prime}}(\Omega_{i})>$,  in the partition $\lq\lq i"$, is straightforward, while the same becomes complicated for $<{\cal Y}_{K \alpha_{i}}(\Omega_{i}) \mid V_{i j}(x_{k}) \mid {\cal Y}_{K^{\prime}  \alpha_{i}^{~\prime}}(\Omega_{i})>$ or $<{\cal Y}_{K \alpha_{i}}(\Omega_{i}) \mid V_{k i} (x_{j})|{\cal Y}_{K^{\prime} \alpha_{i}^{~\prime}}(\Omega_{i})>$ even for Coulomb like central potentials, since $x_{k}$ or $x_{j}$ involves the polar angles $\hat{x_{i}}$ and $\hat{y_{i}}$ and most of the five dimensional integrals have to be 
done numerically. From Eq. (1), we may write \begin{equation}
 \left. \begin{array}{rcl}
\vec{x_{k}} & = & - \cos \sigma_{ki} \vec{x_{i}} + \sin \sigma_{ki}
\vec{y_{i}}\\ 
\vec{y_{k}} & = & - \sin \sigma_{ki} \vec{x_{i}} - \cos \sigma_{ki} \vec{y_{i}}
   \end{array} \right\}
\end{equation}
where $\sigma_{ki}$ = $\tan^{-1} \{(-1)^{P} \sqrt{\frac{M m_{j}}{m_{i} m_{k}}}
\}$, P being odd (even) if ($kij$) is an odd (even) permutation of the
triad (1 2 3).

However, evaluation of the latter matrix elements can be greatly simplified [47]. As the complete sets of HH $\{{\cal Y}_{K \alpha_{i}}(\Omega_{i})\}$, $\{{\cal Y}_{K
\alpha_{j}}(\Omega_{j})\}$ or $\{{\cal Y}_{K \alpha_{k}}(\Omega_{k})\}$ span the same five dimensional angular
hyperspace, any particular member of the given set, say ${\cal Y}_{K \alpha_{i}}(\Omega_{i})$ can be expanded in the complete set of $\{{\cal Y}_{K \alpha_{j}}(\Omega_{j})\}$ through a unitary transformation:
\begin{equation}
{\cal Y}_{K \alpha_{i}}(\Omega_{i})=\sum_{\alpha_{j}} < \alpha_{j} \mid
 \alpha_{i} >_{K L} {\cal Y}_{K \alpha_{j}}(\Omega_{j}) 
\end{equation}
Again, since $K, L, M$ are conserved for Eq. (10) and there is rotational degeneracy 
with respect to the quantum number $M$ for spin independent forces, we have 
\begin{equation}
<\alpha_{j} \mid \alpha_{i}>_{K L} = <l_{x_{j}} l_{y_{j}} \mid l_{x_{i}}
l_{y_{i}}>_{K L}
\end{equation}
Thus, Eq. (10) can be rewritten as 
\begin{equation}
{\cal Y}_{K\alpha_{i }}(\Omega_{i}) = \sum_{l_{x_{j}} l_{y_{j}}}<l_{x_{j}}
l_{y_{j}} \mid  l_{x_{i}} l_{y_{i}}>_{K L} {\cal Y}_{K\alpha_{j}}(\Omega_{j})
\end{equation}
 The M independent coefficients involved in Eq. (11) and (12) are called the Raynal-Revai Coefficients (RRC). Using these coefficients, the matrix element of a central potential $V_{ij}$ becomes 
\begin{equation}
 \begin{array}{rcl}
<{\cal Y}_{K \alpha_{i}}(\Omega_{i}) \mid V_{ij} (x_{k}) \mid {\cal Y}_{K^{\prime} 
 \alpha_{i}^{~\prime}}(\Omega_{i})> & = & \sum_{l_{x_{k}}^{~\prime}
l_{y_{k}}^{~\prime} l_{x_{k}} l_{y_{k}}}<l_{x_{k}} 
l_{y_{k}} \mid  l_{x_{i}} l_{y_{i}}>_{K L}^{*}\\
&\times & <l_{x_{k}}^{~\prime}
l_{y_{k}}^{~\prime} \mid l_{x_{i}}^{~\prime} l_{y_{i}}^{~\prime}
>_{K^{\prime} L} \\  &\times &  <{\cal Y}_{K
\alpha_{k}}(\Omega_{k})\mid V_{ij}(x_{k}) \mid {\cal Y}_{K^{\prime}
 \alpha_{k}^{~\prime}}(\Omega_{k})>
 \end{array}
\end{equation}
 The matrix element on the right side of Eq. (13) resembles the matrix element of $V_{jk}$ in the partition $\lq\lq i$" and can be calculated in a simple manner. Thus, one can calculate matrix element of $V_{ij}$ easily by computing RRC's involved in Eq. (13) using their elaborate expressions from [44-45,48]. Similar treatment can be applied for the calculation of the matrix element of $V_{ki}$. At this point we may also refer the analytical calculation of matrix elements of the effective potential in correlation function HH method by Krivec and Mandelzweig [49].

\section{Results and discussions} 
For the present calculation, we assign the label $\lq i$' to the nucleus of mass $m_i=m_N$ (and charge +Ze), the label $\lq j$' to the negatively charged muon of mass $m_j=m_{\mu}$ (and charge -e) and the label $\lq k$' to the electron of mass $m_k=m_e$ (and charge -e). Hence, for this particular choice of masses, Jacobi coordinates of Eq. (1) in the partition $\lq\lq i$" become 
\begin{equation}
 \left. \begin{array}{rcl}
  \vec{x_{i}} & = &\left[\frac{m_{\mu}m_e(m_{N}+m_{\mu}+m_e)}{m_N(m_{\mu}+m_e)^2} \right]^{\frac{1}{4}} (\vec{r_{j}} - \vec{r_{k}}) \\
  \vec{y_{i}} & = &\left[\frac{m_{\mu}m_e(m_{N}+m_{\mu}+m_e)}{m_N(m_{\mu}+m_e)^2} \right]^{-\frac{1}{4}} (\vec{r_{i}} - \frac{m_{\mu}\vec{r_{j}}+
m_e\vec{r_{k}}}{m_{\mu}+m_e}) 
   \end{array} \right\}
\end{equation}
and the corresponding Schr\"{o}dinger equation (Eq. (7)) is  
\begin{equation}
\begin{array}{lcl}
 \left[-\frac{\hbar^2}{2\mu}\left\{ \frac{d^{2}}{d\rho^{2}}
-\frac{(K+3/2)(K+5/2)}{\rho^{2}}\right\} - E \right] 
U_{K \alpha_{i}}(\rho)&&  \\
+ \sum_{K^{\prime} \alpha_{i}^{~\prime}} <K\alpha_{i}
\mid\frac{\beta_{i}}{\rho 
cos\phi_{i}} - \frac{Z}{\rho\left|\beta_{i}sin \phi_{i}~
\hat{y_{i}}-
\frac{1}{2\beta_{i}} cos\phi_{i}\hat{x_{i}}\right|}&& \\
  -\frac{Z}{\rho\left|
\beta_{i}sin \phi_{i} \hat{y_{i}}+\frac{1}{2\beta_{i}}cos\phi_{i}
\hat{x_{i}}\right|}
             \mid  K^{\prime} \alpha_{i}^{~\prime}
>U_{K^{\prime} \alpha_{i}^{~\prime}}(\rho)& = & 0
\end{array}
\end{equation}
where $\beta_{i}=\left[\frac{m_{\mu}m_e(m_{N}+m_{\mu}+m_e)}{m_N(m_{\mu}+m_e)^2} \right]^{\frac{1}{4}}$ and  $\mu=\left(\frac{m_{N}m_{\mu}m_e}{m_{N}+m_{\mu}+m_e} \right)^{\frac{1}{2}}$ is the effective mass of the system. In atomic units we take $\hbar^{2}=m_e=m=e^{2}=1$. Masses of the particles involved in this work are partly taken from [47,50-52]. Calculation of potential matrix elements of muon-nucleus and electro-nucleus Coulomb interactions $V_{ij}$ and $V_{ki}$ in the partition $\lq\lq i$" are greatly simplified by the use of RRC as discussed in the previous section.

In the ground state of electron-muon three-body system, the total orbital angular momentum, $L$=0 and there is no restriction (on $lx_i$) due to Pauli exclusion principle as electron and muon are non-identical fermions. Since $L=0$, ${l_{x_{i}}= l_{y_{i}}}$, and the set of quantum numbers represented by $\alpha_{i}$ is ${\left\{ l_{x_{i}}, l_{x_{i}}, 0, 0 \right\}}$. Hence, the quantum numbers ${\left\{K\alpha_{i} \right\}}$ can be represented by ${\left\{K l_{x_{i}}\right\}}$ only. Corresponding HH can then be written as
\begin{equation}
 \begin{array}{ccl}
{\cal Y}_{K \alpha_{i}}(\Omega_{i}) & \equiv & {\cal Y}_{K l_{x_{i}}
l_{x_{i}} 0 0}(\Omega_{i}) \\ 
 & =  & ^{(2)}P_{K}^{l_{x_{i}} l_{x_{i}}}(\phi_{i})\left[Y_{l_{x_{i}}
m_{x_{i}}}(\cal Y_{x_{i}},\phi_{x_{i}}) 
Y_{l_{x_{i}} - m_{x_{i}}} (\cal Y_{x_{i}},\phi_{x_{i}})\right]_{0 0} \\
 &  & \left( K \: even \: and \: l_{x_{i}} \: = \: 0, 1, 2, 3, 4, \ldots ,\leq K/2
\right) . 
 \end{array}
\end{equation}

The matrix element of the muon-electron repulsion term in our chosen partition $\lq\lq i$", is 
\begin{equation}
  \begin{array} {rcl}
    <K^{\prime}l_{x_{i}}^{\prime}|\frac{\beta_{i}}{\rho ~cos\phi_{i}}|K
l_{x_{i}}>&=&\frac{\beta_{i}}{\rho} \delta_{l_{x_{i}}^{~\prime}, 
 l_{x_{i}}} \int_{0}^{\pi /2} { ^{(2)}P_{K^{\prime}}}^{l_{x_{i}}
l_{x_{i}}}(\phi) \\
&&\times {^{(2)}P_{K}}^{l_{x_{i}} l_{x_{i}}}(\phi) \sin^{2}\phi ~\cos~\phi ~d\phi\\
  \end{array}
\end{equation}
in which the suffix $i$ on $\phi$ has been dropped deliberately, since $\phi$ is only a variable of integration. Using Eq. (13), matrix elements of the muon-nucleus and electron-nucleus attractive potentials (i.e. second and third terms of the potential in Eq. (16)) in our chosen partition (i.e., partition $\lq\lq i$") respectively become
\begin{equation}
 \begin{array}{rcl}
<K^{\prime} l_{x_{i}}^{\prime}\mid\frac{Z}{r_{ij}}\mid K
l_{x_{i}}> & = & \sum_{l_{x_{k}}}<l_{x_{k}}l_{x_{k}} \mid
l_{x_{i}}^{\prime}l_{x_{i}}^{\prime}>_{K^{\prime}0}^{\*} 
<l_{x_{k}} l_{x_{k}}\mid l_{x_{i}} l_{x_{i}}>_{K  
0} \\
 && <K^{\prime} l_{x_{k}}\mid\frac{Z \beta_{k}}{\rho
cos\phi_{k}}\mid K l_{x_{k}}>. 
 \end{array}
\end{equation}
and
\begin{equation}
 \begin{array}{rcl}
<K^{\prime}l_{x_{i}}^{\prime}\mid\frac{Z}{r_{ik}}\mid K l_{x_{i}}>
&=&\sum_{l_{x_{j}}} <l_{x_{j}}l_{x_{j}}\mid 
 l_{x_{i}}^{~\prime}l_{x_{i}}^{\prime}>_{K^{\prime}0}^{\*}
<l_{x_{j}}l_{x_{j}}\mid 
l_{x_{i}}l_{x_{i}}>_{K 0} \\ 
 &&<K^{\prime}l_{x_{j}}\mid\frac{Z\beta_{j}}{\rho
cos\phi_{j}}\mid K l_{x_{j}}>.  
 \end{array}
\end{equation}
Sums over $l_{x_{k}}^{\prime}$ and
$l_{x_{j}}^{\prime}$ respectively in Eq. (18) and (19) have been performed using the Kronecker
 - $\delta$'s (as in Eq. (17)). Thus the evaluation of the matrix elements of $\underline{all}$ the
potential components become practically simple and easy to handle them numerically.

One of the major drawbacks of HH expansion method is its slow rate of convergence for Coulomb-type long range interaction potentials, unlike for the Yukawa-type short-range potentials for which the convergence is reasonably fast [46,53]. Hence, to achieve the desired degree of convergence, sufficiently large $K_m$ value has to be included in the calculation. But, if all $K$ values up to a maximum of $K_m$ are included in the HH expansion then the number of the basis states can be determined by relation \begin{equation}  N_{K_m}= \frac{(K_m+2)(K_m+4)}{8} \end{equation}
It follows from Eq. (20) that number of basis states and hence the size of coupled differential equations (CDE) (Eq. (7)) increases rapidly with increase in $K_m$. For instance, one has to solve 561 CDEs for $K_m = 64$ which leads the calculation towards instability. For the available computer facilities, we are allowed to solve up to $K_m=28$ reliably. Energies for still higher $K_m$ are obtained following extrapolation scheme of Schneider [54] discussed in our previous work [47]. The calculated ground state energies ($B_{K_m}$) with increasing $K_m$ for muonic helium ($^{\infty}$He$^{2+}\mu^-e^-$), muonic lithium  ($^{\infty}$Li$^{3+}\mu^-e^-$) and muonic berilium ($^{\infty}$Be$^{4+}\mu^-e^-$) are presented in columns 2, 4 and 6 of Table I. Energies for a number of muonic atom/ions of different atomic number (Z) at $K_m=28$ are presented in column 3 of Table II. The extrapolated energies for few of the above systems are presented in bold in column 4 of Table II.

The pattern of convergence of the energy of the lowest bound S-state with respect to increasing $K_m$ can be checked by gradually increasing $K_m$ values in suitable steps ($dK$) and comparing the relative energy difference $\eta =\frac{B_{K+dK}-B_K}{B_{K+dK}}$ with that found in the previous step. From Table I, it can be seen that at $K_m=28$, the energy of the lowest bound S-state of ${e^-\mu^-}$ $^{\infty}$He$^{2+}$ converges up to 3rd decimal places and similar convergence trends are observed in the remaining cases.

Furthermore, although an easy computation of the matrix element of ${\frac{1}{r_{ij}}}$ in the partition $\lq\lq i$" is possible by the method of Ref.[61 of 47], it is not so easy for potentials other than Coulomb or harmonic type. For an arbitrary shape of interaction potential, a direct computation of the matrix element of the potential will involve five dimensional angular integrations which lead the calculation very time consuming and leaves windows open for inaccuracies to creep in easily. Thus for accurate and faster computation of energy, role of RRC in HH method is unique and essential.
%\newpage
\begin{figure}
\centering
\fbox{\includegraphics[width=0.75\linewidth, height=0.6\linewidth]{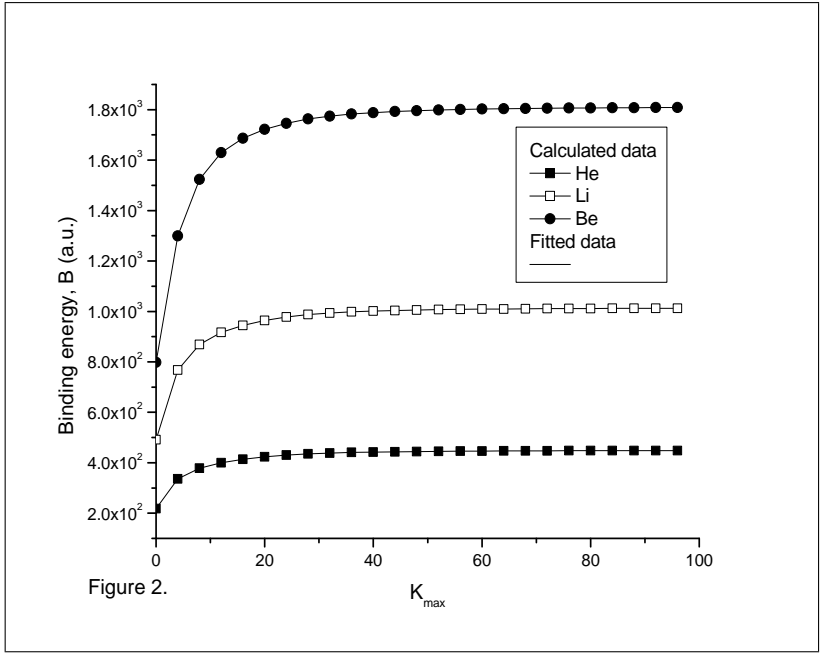}}
\caption{Pattern of dependence of the ground-state energy (B) of muonic atom/ions on the increase in $K_{max}$.}
\label{fig:boxed_graphic}
\end{figure}

\begin{figure}
\centering
\fbox{\includegraphics[width=0.75\linewidth, height=0.6\linewidth]{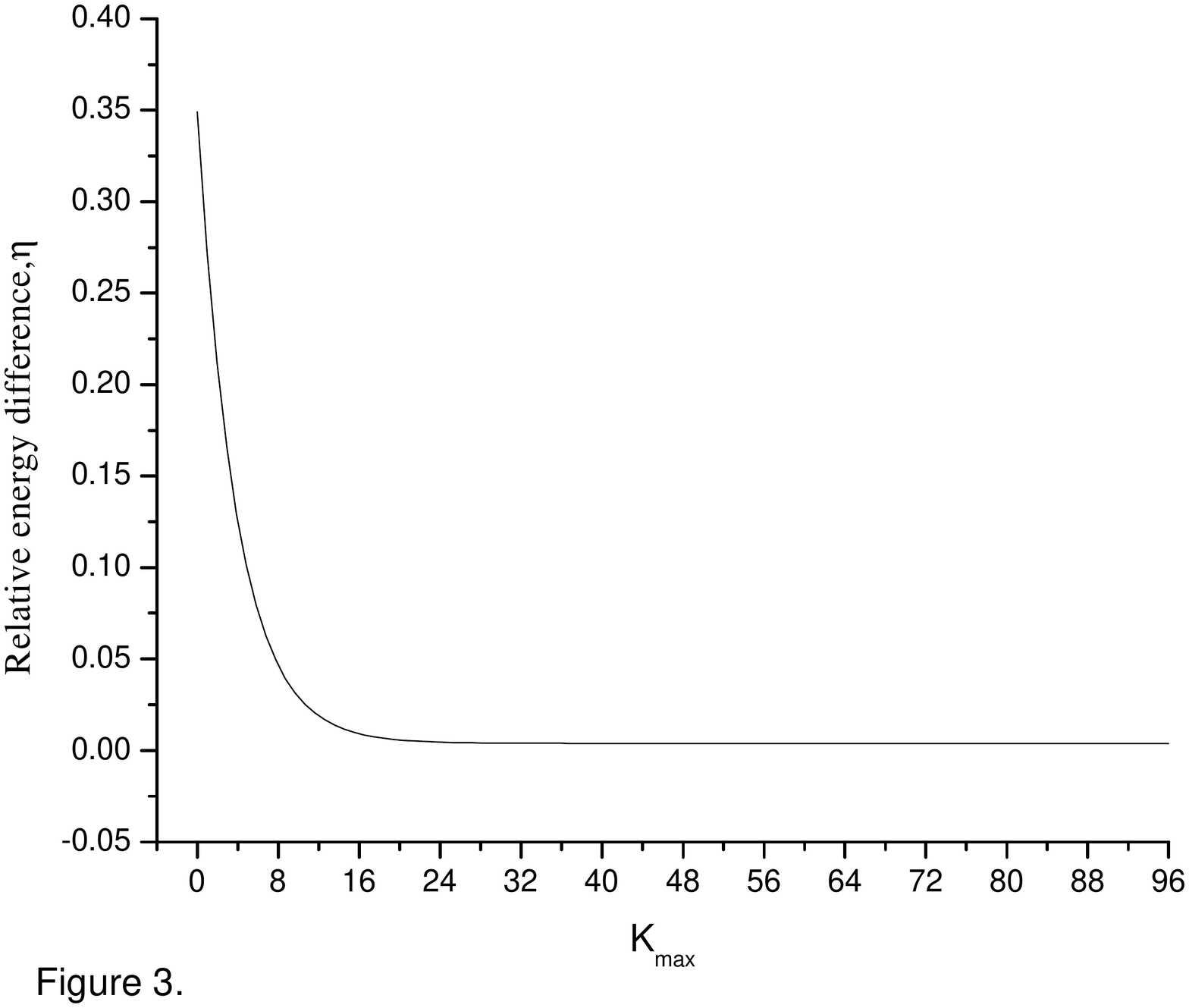}}
\caption{Pattern dependence of the ground-state relative energy difference $\eta=\frac{B_{K_m+4}-B_{K_m}}{B_{K_m+4}}$ of muonic helium ($^{\infty}$He$^{2+}\mu^-e^-$) on the increase in $K_{max}$.}
\label{fig:boxed_graphic}
\end{figure}

\begin{figure}
\centering
\fbox{\includegraphics[width=0.75\linewidth, height=0.6\linewidth]{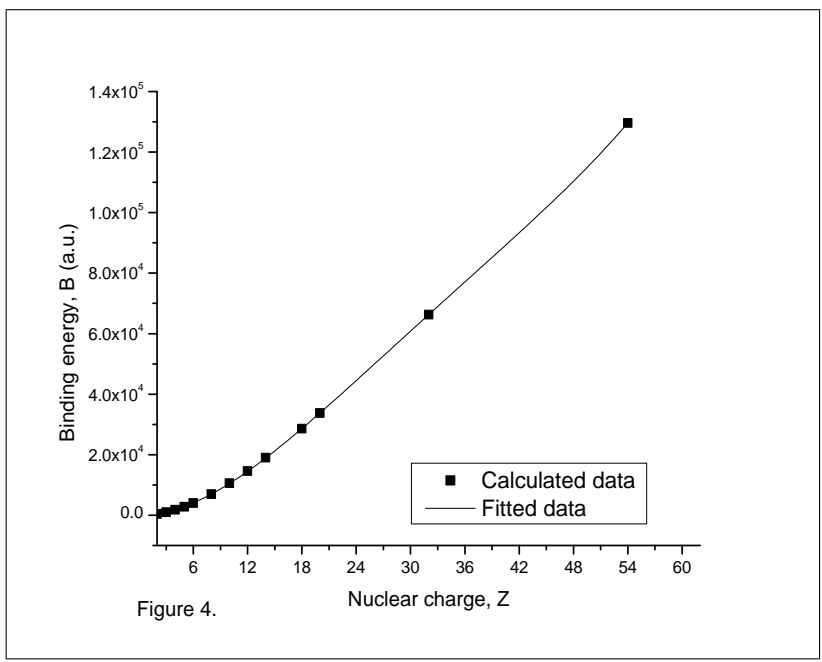}}
\caption{Pattern of dependence of the ground-state energy (B) of muonic atom/ions on the increase in nuclear charge Z.}
\label{fig:boxed_graphic}
\end{figure}

 The pattern of increase in binding energy (B) with respect to increasing $K_{max}$ is shown in Figure 3 for few representative cases. In Figure 4 the relative energy difference $\eta$ is plotted against $K_{max}$ to demonstrate the relative convergence trend in energy. The calculated ground state energies muonic three-body systems of different nuclear charge Z (and of infinite nuclear mass), have been plotted against Z as shown in Figure 4 to study the dependence of the bound state energies on the strength of the nuclear charge using data from Table II. The curve of Figure 4 shows a gradual increase in energy with the increase in the strength nuclear charge Z approximately following the empirical equation
\begin{eqnarray}    B(Z)&=&-154.64851+4.47088 Z+131.84786Z^2-2.79311 Z^3+0.02174Z^4  \end{eqnarray}
Eq. (21) may be used to estimate the ground state energy of muonic atom/ions of given Z having infinite nuclear mass. Finally, in Table II, energies of the lowest bound S-state of several muonic three-body systems obtained by numerical solution of the coupled differential equations by the renormalized Numerov method [55] have been compared with the ones of the literature wherever available. Since reference values are not available for systems having nuclear charge $Z>3$, we made a crude estimation of the ground-state ($1s_e1s_{\mu}$) energies following the relation \begin{equation} B_{est} = \frac{1836AZ^2}{2}\left[\frac{1}{1+1836A}+\frac{206.762828}{206.762828+1836A}\right](a.u.)\end{equation} where A is the mass number of the nucleus. Here we assumed two hydrogen-like subsystems for the muonic atom/ions. 

\section{Conclusion}
In conclusion, we note that the calculated ground-state energy of muonic helium and muonic lithium at $K_m=28$ listed in column 2 of Table II are greater than the corresponding reference values listed in clolumn 4 of Table II. This dicrepancies may have arose due to the CUSP condtions applicable to Coulomb systems which have not been accommodated in the present calculations. It can be seen that the estimated energies ($B_{est}$) in column 2 of Table II are less than the corresponding calculated energies ($B_{calc}$) in column 3 of Table II for systems having $Z\leq 10$ while the same for $Z>10$ becomes larger than $B_{calc}$. The facts indicate a weaker correlation betwen electron-muon pair in systems having nuclear charge $Z<10$ while a stronger correlation in systems having $Z>10$. The fully converged extrapolated energies ($B_{conv}$) in column 4 of Table II, obtained for a very large $K_{max}$ following prescription of [47] are larger than the estimated values ($B_{est}$) in all cases. Which indicates that the electron-muon repulsion term plays a vital role in the binding mechanism of the electron-muon-nucleus Coulomb three-body systems. It may also be noted that the RRC's being independent of r, needs to be calculated once only and stored, resulting in an efficient and highly economical numerical computation. Finally, it may be mentioned here that in the cases of higly charged muonic ions relativistic correction term has to be introduced to get the best results.\\
\hspace*{1cm} The author acknowledges Aliah University for providing computer facilities.

\section{Figure Caption}
\vspace{0.1cm}
\begin{list}{•}{•}
\item Fig. 1. Choice of Jacobi coordinates in different partitions of a three-body system.
\item Fig. 2. Pattern of dependence of the ground-state energy (B) of muonic atom/ions on the increase in $K_{max}$.
\item Fig. 3. Pattern dependence of the ground-state relative energy difference $\eta=\frac{B_{K_m+4}-B_{K_m}}{B_{K_m+4}}$ of muonic helium ($^{\infty}$He$^{2+}\mu^-e^-$) on the increase in $K_{max}$.
\item Fig. 4. Pattern of dependence of the ground-state energy (B) of muonic atom/ions on the increase in nuclear charge Z.
\end{list}
\section{Tables}
\newpage
\begin{table}
\begin{center}
\begin{scriptsize}
{\bf Table I. Energy (B) of the lowest bound S-state of electron-muon three-body systems at different $K_{max}$ along with the corresponding relative energy difference $\eta$.}\\
\vspace{5pt}
\begin{tabular}{cc cc cc c}\hline
&\multicolumn{6}{c}{Binding energies (B) and corresponding relative energy difference ($\eta$)}\\
\cline{2-7}
System&\multicolumn{2}{c}{$^{\infty}$He$^{2+}\mu^-e^-$}&\multicolumn{2}{c}{$^{\infty}$Li$^{3+}\mu^-e^-$}&\multicolumn{2}{c}{$^{\infty}$Be$^{4+}\mu^-e^-$}\\
$K_{max}$&$B$&$\eta$&$B$&$\eta$&$B$&$\eta$\\\hline
0&217.78577&0.352207&490.78306&0.360783&798.35396&0.385798\\
4&336.19645&0.110907&767.78745&0.116652&1299.82240&0.146813\\
8&378.13435&0.055125&869.17847&0.052009&1523.49147&0.065150\\
12&400.19520&0.033222&916.86372&0.029708&1629.66350&0.033758\\
16&413.94723&0.022228&944.93547&0.019627&1686.59904&0.020329\\
20&423.35753&0.015913&963.85254&0.014046&1721.59813&0.013800\\
24&430.20330&0.011949&977.58415&0.010573&1745.68917&0.010153\\
28&435.40598&0.006945&988.03084&0.005984&1763.59496&0.006227\\
32&438.45105&0.005019&993.97864&0.004280&1774.64585&0.004433\\
36&440.66258&0.003718&998.25075&0.003143&1782.54755&0.003242\\
40&442.30725&0.002814&1001.39787&0.002360&1788.34611&0.002427\\
44&443.55559&0.002170&1003.76711&0.001807&1792.69697&0.001853\\
48&444.52011&0.001700&1005.58464&0.001408&1796.02503&0.001439\\
52&445.27709&0.001351&1007.00210&0.001113&1798.61395&0.001135\\
56&445.87945&0.001087&1008.12376&0.000891&1800.65799&0.000907\\
60&446.36474&0.000885&1009.02291&0.000722&1802.29325&0.000734\\
64&446.76007&0.000728&1009.75208&0.000591&1803.61700&0.000600\\
68&447.08534& 0.000604&1010.34962&0.000489&1804.70001&0.000495\\
72&447.35542&0.000505&1010.84392&0.000408&1805.59460&0.000413\\
76&447.58153&0.000426&1011.25636&0.000343&1806.34002&0.000346\\
80&447.77226&0.000362&1011.60320&0.000290&1806.96611&0.000293\\\hline
\end{tabular}
\end{scriptsize}
\end{center}
\end{table}

\begin{table}
\begin{center}
\begin{scriptsize}
{\bf Table II. Energy (B) of the lowest bound S-state of electron-muon-nucleus three-body systems.}\\
\vspace{5pt}
\begin{tabular}{lrcrl}\hline
System&\multicolumn{4}{c}{Binding energies expressed in atomic unit (a.u.)}\\
\cline{2-5}
& Estimated [Eq.(22)]&\multicolumn{2}{c}{Present Calculation}&Other Results\\
\cline{3-4}
&$B_{est}$& $B_{K_m=28}$ & $B_{conv}$ &\\\hline
$e^-\mu^-$$^3$He$^{2+}$&400.574&420.424&{\bf 433.870}&399.042$^a$, 399.043$^b$\\ 
$e^-\mu^-$$^4$He$^{2+}$&404.212&424.017&{\bf 437.538}&402.637$^c$, 402.641$^d$\\
$e^-\mu^-$$^{\infty}$He$^{2+}$&415.537&435.406&{\bf 449.160}&414.036$^e$, 414.037$^f$\\

$e^-\mu^-$$^6$Li$^{3+}$&917.814&970.737&{\bf 996.404}&915.231$^e$, 915.231$^g$\\
$e^-\mu^-$$^7$Li$^{3+}$&920.224&973.166&{\bf 998.886}&917.649$^e$, 917.650$^g$\\
$e^-\mu^-$$^{\infty}$Li$^{3+}$&934.957&988.031&{\bf 1014.080}&932.457$^e$\\

$e^-\mu^-$$^{9}$Be$^{4+}$& 1641.703&1745.087 & & \\
$e^-\mu^-$$^{\infty}$Be$^{4+}$&1662.146&1763.595 &{\bf 1811.404}  &\\

$e^-\mu^-$$^{10}$B$^{5+}$ &2568.319&2732.374 & &\\
$e^-\mu^-$$^{\infty}$B$^{5+}$&2597.104 &2761.200&{\bf 2854.793} &\\

$e^-\mu^-$$^{12}$C$^{6+}$&3705.224 &3937.535 &{\bf }&\\ 
$e^-\mu^-$$^{\infty}$C$^{6+}$&3739.829&3971.528 &{\bf 4161.585}  &\\

$e^-\mu^-$$^{16}$O$^{8+}$&6602.337&6907.068 &{\bf }&\\ 
$e^-\mu^-$$^{\infty}$O$^{8+}$&6648.585&6949.141 &{\bf 7598.178} &\\

$e^-\mu^-$$^{20}$Ne$^{10+}$&10330.524 &10486.654 &{\bf }&\\ 
$e^-\mu^-$$^{\infty}$Ne$^{10+}$&10388.414&10534.362 &{\bf 12176.554}&\\											

$e^-\mu^-$$^{24}$Mg$^{12+}$&14889.783 &14539.329 &{\bf }&\\
$e^-\mu^-$$^{\infty}$Mg$^{12+}$&14959.316&14590.826&{\bf 17957.871}&\\

$e^-\mu^-$$^{28}$Si$^{14+}$&20280.115 &18956.238 &{\bf }&\\
$e^-\mu^-$$^{\infty}$Si$^{14+}$&20361.291& 19010.171&{\bf 24997.977}&\\

$e^-\mu^-$$^{32}$S$^{16+}$&26501.520&23653.644&{\bf}&\\
$e^-\mu^-$$^{\infty}$S$^{16+}$&26594.340&23709.055&{\bf 33343.016}&\\

$e^-\mu^-$$^{40}$Ar$^{18+}$&33564.416&28574.033&{\bf}&\\
$e^-\mu^-$$^{\infty}$Ar$^{18+}$&33658.461&28624.646&{\bf 43031.333}&\\

$e^-\mu^-$$^{40}$Ca$^{20+}$&41437.550&33152.978&{\bf }&\\
$e^-\mu^-$$^{\infty}$Ca$^{20+}$&41553.656&33709.888&{\bf 54041.928}&\\

$e^-\mu^-$$^{73}$Ge$^{32+}$&106214.286&66247.363 &{\bf }& \\
$e^-\mu^-$$^{\infty}$Ge$^{32+}$&106377.359&66295.584 &&\\

$e^-\mu^-$$^{132}$Xe$^{54+}$&302669.161&129587.296&{\bf }&\\
$e^-\mu^-$$^{\infty}$Xe$^{54+}$&302926.152&129628.491&&\\

$e^-\mu^-$$^{222}$Rn$^{86+}$&767939.397&223901.011&{\bf }&\\
$e^-\mu^-$$^{\infty}$Rn$^{86+}$&768327.099&223936.847 &&\\

$e^-\mu^-$$^{238}$U$^{92+}$&878861.487&241689.709&{\bf }&\\
$e^-\mu^-$$^{\infty}$U$^{92+}$&879275.361&241725.061&&\\\hline
\end{tabular}\\
$^a$Ref[2,16,18,56], $^b$Ref[57], $^c$Ref[2-3,16,18,57-58], $^d$Ref[1,34,59], 
$^e$Ref[2], $^f$Ref[18], $^g$Ref[29]\\
\end{scriptsize}
\end{center}
\end{table}

\end{document}